Original article

# Rise Time Effects of a Portable Inductive Energy Storage Pulse Generator on NO Production in Spark Discharges

*[1]Cheng-Hsiao Hsueh, [1]Yu-Hsuan Chen, [1]Chao-Yu Chen, [1,2]Yun-Chien Cheng*

[1]Department of Mechanical Engineering, National Yang Ming Chiao Tung University, Hsinchu 30010, Taiwan
[2] Department of Electrical Engineering, National Taiwan University, Taipei, Taiwan

*Address all correspondence to: Yun-Chien Cheng, Department of Electrical Engineering, National Taiwan University, Taipei, Taiwan; Tel.: +(886) 976114671; E-mail: yccheng@gmail.com

**Abstract**

This study investigates pulse waveform adjustment in a portable inductive energy storage (IES) pulsed power systems and its effect on nitric oxide (NO) generation in atmospheric spark discharges. While previous studies have focused on pulse energy, width, and repetition rate, the role of rise time remains insufficiently explored. Furthermore, existing pulse generators with adjustable rise time capabilities are bulky and unsuitable for portable applications. To address this, this study investigates various rise time control strategies within compact IES architecture and employs a parallel high voltage capacitor to generate adjustable waveforms.

Optical emission spectroscopy was used to measure excited $N_2$ and NOγ emissions, and NO analyzer monitored NO concentration. Besides, the electron excitation temperature, plasma-absorbed power, and rotational temperature were estimated to clarify the influence of waveform characteristics on plasma chemistry. Results show that although the rise time can be adjusted, it is coupled with the pulse width and input energy in system. Comparative analysis of spectral intensities, NO concentration, electron excitation temperature, and absorption energy reveal that NO production depends primarily on energy input rather than rise time, and electrode temperature varies minimally under different waveform conditions, suggesting that NO formation is primarily influenced by electron impact and ionization processes.

This work proposes a portable, simplified IES-based high voltage pulsed power generator, identifies the constraints associated with rise time control, and demonstrates the feasibility of waveform adjustment. The findings clarify the dominant mechanisms governing NO generation and provide a reference for the design of future portable plasma-based NO production systems.

Keywords: Inductive Energy Storage (IES), High voltage pulsed spark discharge, Pulse waveform adjustment, Portable pulse generator

# I. Introduction

Nitric oxide (NO) plays a critical role in physiological regulation and is involved in vasodilation, neurotransmission, and immune modulation [1]. Recent studies have further demonstrated its potential in clinical applications such as inhalation therapy [2,3], wound healing [4], and cancer therapy [5]. Because the therapeutic efficacy of NO is highly concentration dependent, different diseases require specific dosage ranges. Consequently, the development of lightweight, safe, and controllable NO generation and delivery technologies has become an important research focus.

Currently, clinical NO is mainly supplied from high pressure gas cylinders and delivered through complex dilution and delivery systems. These systems are costly, bulky, and difficult to operate, restricting NO therapy primarily to short-term, in-hospital use and limiting its applicability to acute or long term treatments [6].

To overcome these limitations, plasma-based NO generation devices have attracted increasing attention owing to their potential for low cost, portable, and on-site operation [7]. In atmospheric plasmas, NO is mainly produced through electron impact reactions that generate reactive nitrogen and oxygen species [7–9], and through thermal dissociation followed by the Zeldovich mechanism at sufficiently high temperatures [9–11]. Elevated temperatures also promote the decomposition of $O_3$ and $NO_2$, thereby suppressing NO oxidation and facilitating NO accumulation [12,13]. Although trace amounts of $NO_2$ and $O_3$ may be produced, previous studies have shown that appropriate filtration can effectively remove these byproducts and enable clinically safe NO delivery without adverse biological effects [6,14]. These results indicate the strong potential of plasma-based NO generation for clinical translation.

Among various discharge modes, high voltage pulsed arc discharges can generate high temperature, high density plasmas within short durations and have been shown to enhance NO production while reducing $NO_2$ and $O_3$ yields [9,15,16]. Prior studies have demonstrated that input energy [17,18], pulse width [19], and repetition rate [8,17,20,21] significantly affect NO yield. However, the influence of pulse rise time in arc discharges has received little attention. In other discharge types, shorter rise times have been reported to increase electron energy and enhance $N_2^*$ and $N_2^+$ populations [22,23], suggesting that rise time may represent an important but implicit parameter for NO generation, whereas its actual role in arc-related discharges remains insufficiently understood.

Another key challenge for portable NO generators is the pulsed power supply. Existing rise-time-adjustable high voltage sources typically rely on bulky magnetic components or complex control architectures [24–27], limiting portability. Inductive energy storage (IES) circuits offer a compact alternative with simple topology and low component count [28,29]. In an IES system, energy is first stored in an inductor or transformer and then rapidly released through a high speed switching device to generate high voltage pulses. However, in compact IES architectures, changing rise time is strongly constrained and often coupled with other waveform parameters, and systematic investigations remain scarce.

In practical implementations, discharge mode selection must also be considered. Although pulsed arc discharges provide high NO productivity, sustained conductive channels may cause electrode erosion and thermal damage, whereas transient spark discharges retain similar reaction characteristics while reducing average energy input and electrode thermal stress [9,21]. Accordingly, while pulsed arc discharge studies are adopted as the primary theoretical reference, transient spark discharges are selected for experimental implementation in this work.

Motivated by the above considerations, this study develops a portable IES pulsed generator and systematically compares multiple rise time adjustment approaches to identify a feasible and predictably controllable method. The resulting waveforms are applied to spark discharges, and their effects on NO generation are investigated using optical emission spectroscopy, $NO_x$ and $O_3$ measurements, electron excitation temperature estimation, power analysis, and temperature diagnostics.

## II. Materials and Methods

### A. IES Circuit Architecture and Pulse Rise Time Change Methods

(a)

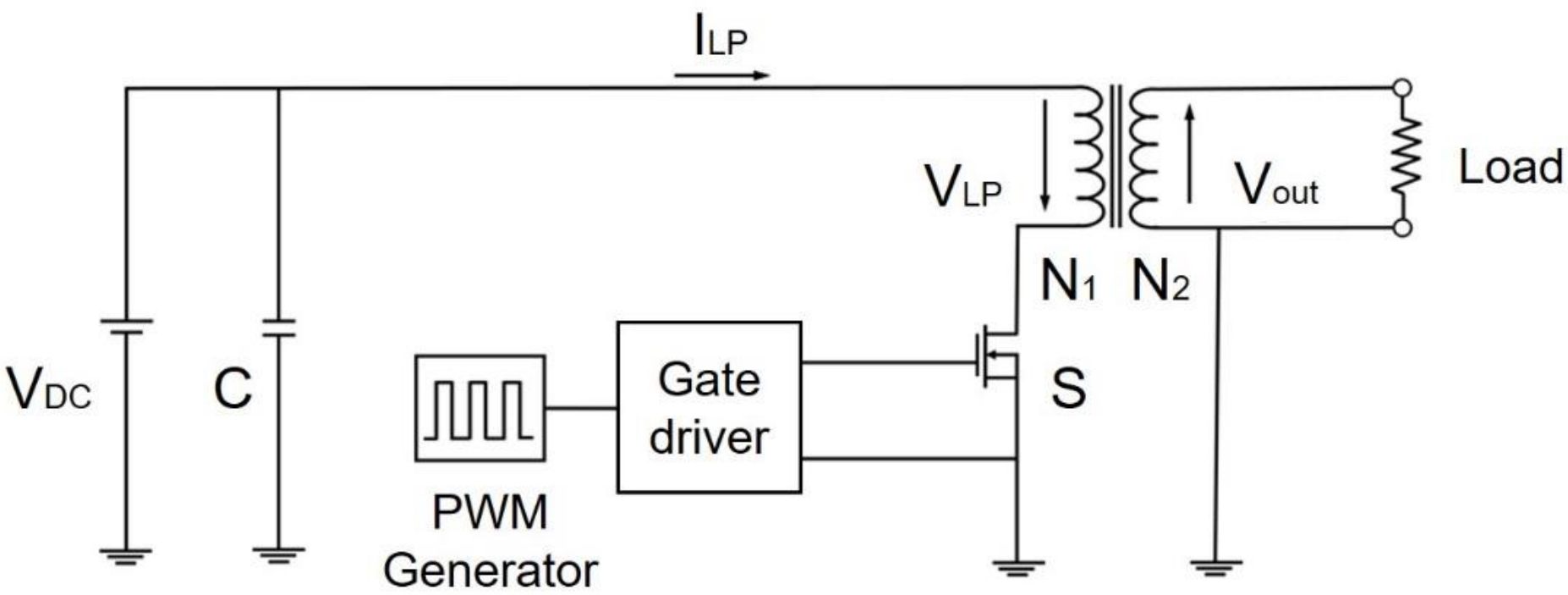

(b)

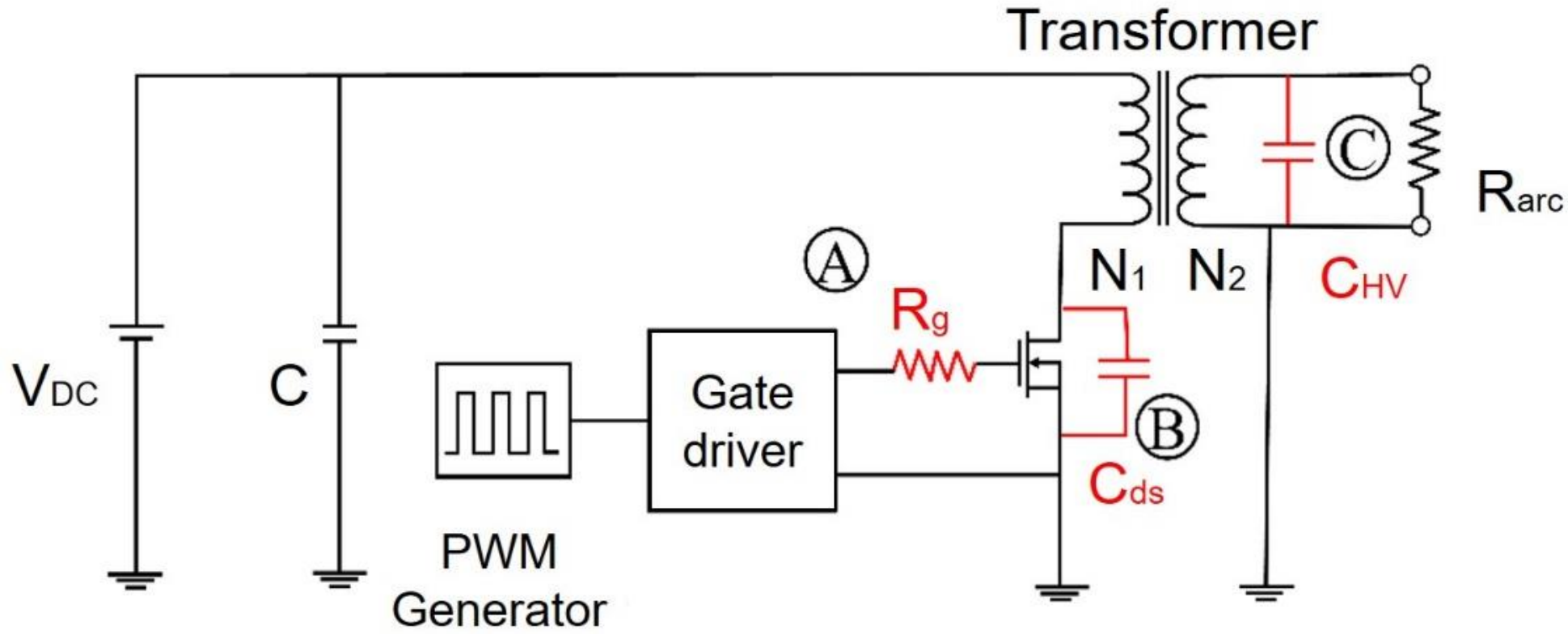

Fig. 1. (a) IES circuit architecture. (b) Pulse rise time change methods using (A) $R_g$, (B) $C_{ds}$, and (C) $C_{HV}$.

Fig. 1(a) shows the base architecture of the IES high voltage pulsed generator, which is based on a SiC MOSFET–transformer configuration. A DC power supply (GPD-3303S, GW Instek) with a 22 µF capacitor provides current to the transformer primary. The SiC MOSFET (AIMW120R060M1H, Infineon) is driven by a gate driver (IXDN604PI, IXYS) using a PWM signal generated by a controller (XIAO SAMD21, Seeed Studio) at 5 kHz, 5% duty cycle, and 15 V amplitude. The transformer employs an EPC-25 ferrite core with primary and secondary windings of 6 and 128 turns, corresponding to inductances of 15 µH and 6.3 mH, respectively.

When the MOSFET is switched, the interruption of the primary current induces a voltage across the primary inductor according to Faraday's law:

$$V_{Lp}(t) = L_{Lp}\frac{dI_{Lp}(t)}{dt} \tag{1}$$

The induced primary voltage is stepped up by the transformer, producing the secondary output voltage:

$$V_{out} = V_{load} = V_{Lp}(t)\frac{N_2}{N_1} \tag{2}$$

To investigate pulse rise time adjustment in the IES circuit, three approaches are employed, as illustrated in Fig. 1(b).

Method (A) adjusts the external MOSFET gate resistance $R_g$ to control the gate current during the Miller plateau. A larger $R_g$ limits the gate driving current $I_G$, slowing the fall of the drain–source voltage $V_{DS}(t)$ during turn on. The instantaneous primary voltage can be expressed as:

$$V_{Lp}(t) = V_{DC} - V_{DS}(t) \tag{3}$$

By differentiating Eq. (3) with respect to time, we obtain：

$$\frac{dV_{Lp}(t)}{dt} = -\frac{dV_{DS}(t)}{dt} \tag{4}$$

Therefore, a slower decrease in $V_{DS}(t)$ directly results in a reduced rate of change of $V_{Lp}(t)$, leading to a longer rise time of the secondary output voltage.

Method (B) introduces an external drain–source capacitor $C_{ds}$ to increase the equivalent MOSFET output capacitance $C_{oss}$, defined as:

$$C_{OSS} = C_{GD} + C_{DS} + C_{ds} \tag{5}$$

During MOSFET turn off, the primary inductive current charges $C_{oss}$, causing $V_{DS}(t)$ to rise. Under a simplified approximation:

$$I_{Lp} \approx C_{OSS}\frac{dV_{DS}(t)}{dt} \tag{6}$$

An increased $C_{oss}$ therefore reduces $dV_{DS}$ / dt, resulting in a slower rise of the induced secondary voltage.

Method (C) connects a high-voltage capacitor $C_{HV}$ in parallel with the transformer secondary. This concept is similar to that reported in [30]. At the initial stage of the voltage rise, the secondary current primarily charges $C_{HV}$, and the output voltage rising rate can be approximated as:

$$I_{Ls}(t) = C_{HV}\frac{dV_{out}(t)}{dt} \tag{7}$$

Thus, a larger $C_{HV}$ directly reduces the rate of change of the output voltage and produces a gentler rising edge. In addition, the capacitor absorbs part of the high frequency energy, suppressing voltage spikes.

Based on these mechanisms, seven values of $R_g$ (30, 90, 210, 330, 510, 590, and 1000 Ω), five values of $C_{ds}$ (10 pF, 100 pF, 1 nF, 10 nF, and 100 nF), and four values of $C_{HV}$ (0 pF, 10 pF, 22 pF, and 33 pF) were selected for experimental evaluation. For Methods (B) and (C), the gate resistance was fixed at 90 Ω.

## B. Experimental Setup for Pulsed Spark Discharge

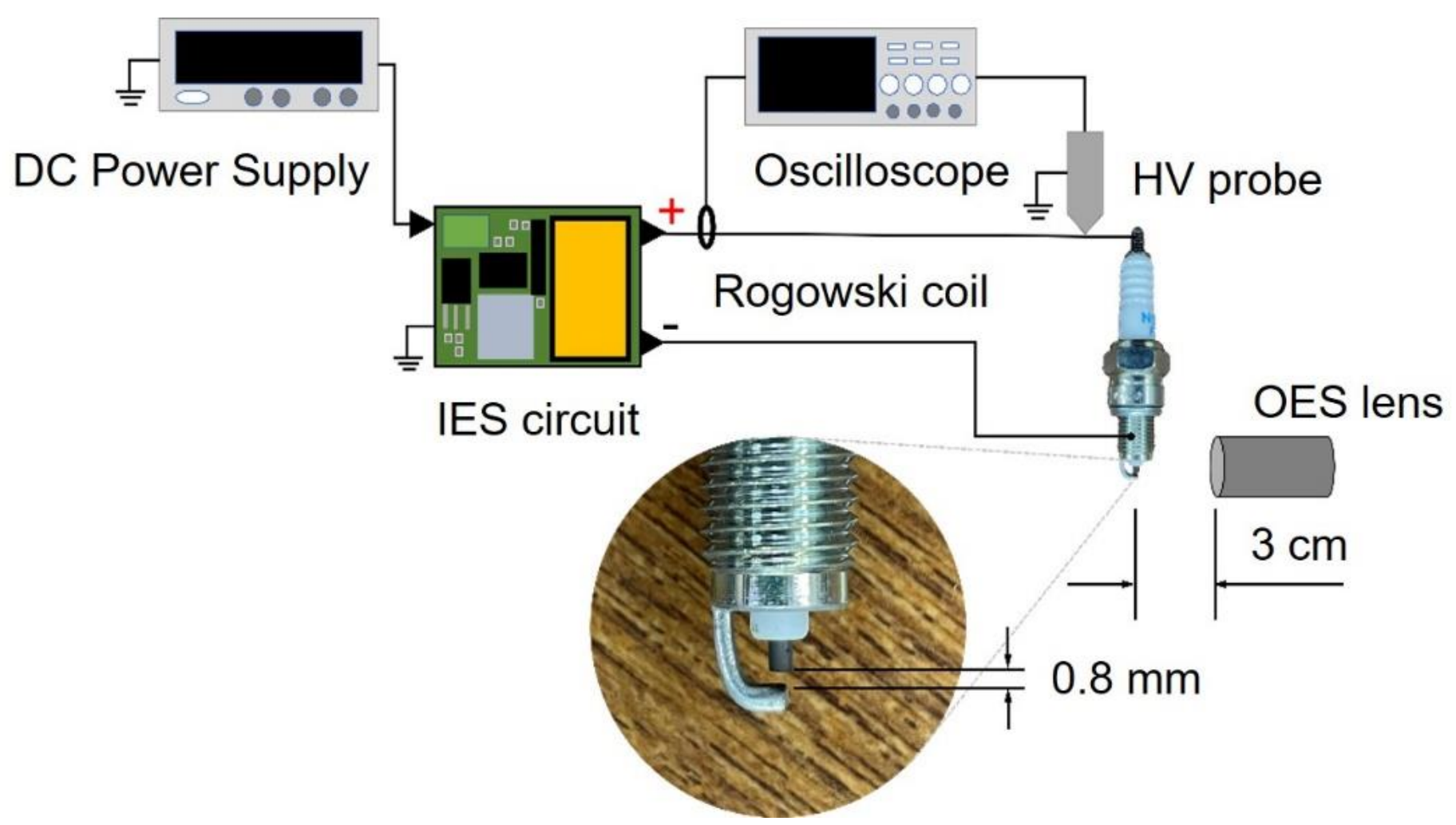

Fig. 2. Experimental setup for pulsed spark discharge.

Fig. 2 shows the experimental setup for pulsed spark discharge. A DC power supply (GPD-3303S, GW Instek, Taiwan) supplied the energy storage current to the transformer primary. The output voltage and current were measured using a high voltage probe (P6015A, Tektronix, United States) and a current probe (CM-100-MG, Ion Physics Corporation, United States), respectively, and recorded with an oscilloscope (InfiniiVision DSOX3024T, Keysight, United States).

A spark plug (LR7D, NGK Spark Plug Corporation, Japan) was used as the discharge electrode to achieve both circuit miniaturization and stable, repeatable discharges. Measurements were performed within 0–3 min after ignition, and the system was allowed to rest for 10 min between experiments to return to its initial state.

### C. Measurement Methods and Instrumentation

Optical emission from the discharge region was measured using a spectrometer (SR2 XR, Ocean Insight, United Kingdom) to examine the influence of different rise time waveforms on NO-related species. The collection lens was fixed 3 cm from the electrode center for all measurements to ensure consistent observation geometry.

**Electron Excitation Temperature Estimation**

The electron excitation temperature was estimated using the line intensity ratio method [31]. Assuming a Boltzmann distribution of excited state populations, the emission intensity of a transition n→m is given by

$$I_{nm} = \frac{hc}{4\pi}\frac{g_n}{g_0}\frac{A_{nm}}{\lambda_{nm}}N_0\exp(\frac{E_n}{K_B T_e}) \tag{8}$$

Where $I_{nm}$ and $\lambda_{nm}$ are the measured emission intensity and wavelength; h is Planck's constant; c is the speed of light; $g_n$ and $g_0$ are the statistical weights of the upper and ground states; $A_{nm}$ is the transition probability; $N_0$ is the ground state number density; $E_n$ is the excitation energy; $K_B$ is the Boltzmann constant; and $T_e$ is the electron excitation temperature.

For two spectral lines from the same ionization state, $T_e$ can be obtained from

$$\frac{I_{nm}}{I_{qp}} = \frac{g_n A_{nm} \lambda_{qp}}{g_q A_{qp} \lambda_{nm}}\exp(-\frac{E_{nm}-E_{qp}}{K_B T_e}) \tag{9}$$

High purity argon (99.99%; Chien-Fa Gas Co., Taiwan) was introduced at 200 sccm approximately 1.3 cm from the discharge gap using a flow meter (Yung Hsin Co., Taiwan) to provide Ar I emission lines for analysis. This flow rate was verified not to affect the spark plug discharge.

**Plasma Absorbed Power Estimation**

The plasma absorbed power was estimated following the method in [32]. The voltage measured at the spark plug includes both the discharge-gap voltage and the voltage drop across the internal resistance of the spark plug. Therefore, the actual gap voltage is

$$V_g(t) = V(t) - I(t)R_p \tag{10}$$

Where V(t) and I(t) are the instantaneous voltage and current measured at the spark plug electrodes, and $R_p$ is the equivalent internal resistance. The energy delivered to the discharge gap is then

$$E_{gap} = \int_{t0}^{t1} (V(t) * I(t) - I(t)^2 R_p)\, dt \tag{11}$$

Where $t_0$ and $t_1$ denote the start and end times of the discharge., respectively. The plasma absorbed power is obtained by multiplying $E_{gap}$ by the pulse repetition rate (PRR):

$$P_{gap} = \int_{t0}^{t1} (V(t) * I(t) - I(t)^2 R_p)\, dt * PRR \tag{12}$$

**NOx and O3 Gas Measurement Setup**

Fig. 3 shows a schematic of the gas measurement setup. A polystyrene (PS) bottle was used as the gas chamber to minimize environmental influence. An air coupler with a 6 mm opening served as the inlet, and a coupler with an 8 to 6 mm reduction served as the outlet. All ports and connectors were sealed with hot melt adhesive and sealing tape to prevent gas leakage.

Air was supplied by an air pump (ACO-9610, Hailipai, China), and a flow meter (Shin-Zen Instruments, Taiwan) provided a purge flow of 1 slm toward the outlet. The post-discharge gas composition was measured using nitric oxide (EST-1012), ozone (EST-1015H), and nitrogen dioxide (EST-1013) analyzers (Environmental Sensor Technology, United States). Prior tests confirmed that the analyzer connection order had no significant effect on the measured concentrations.

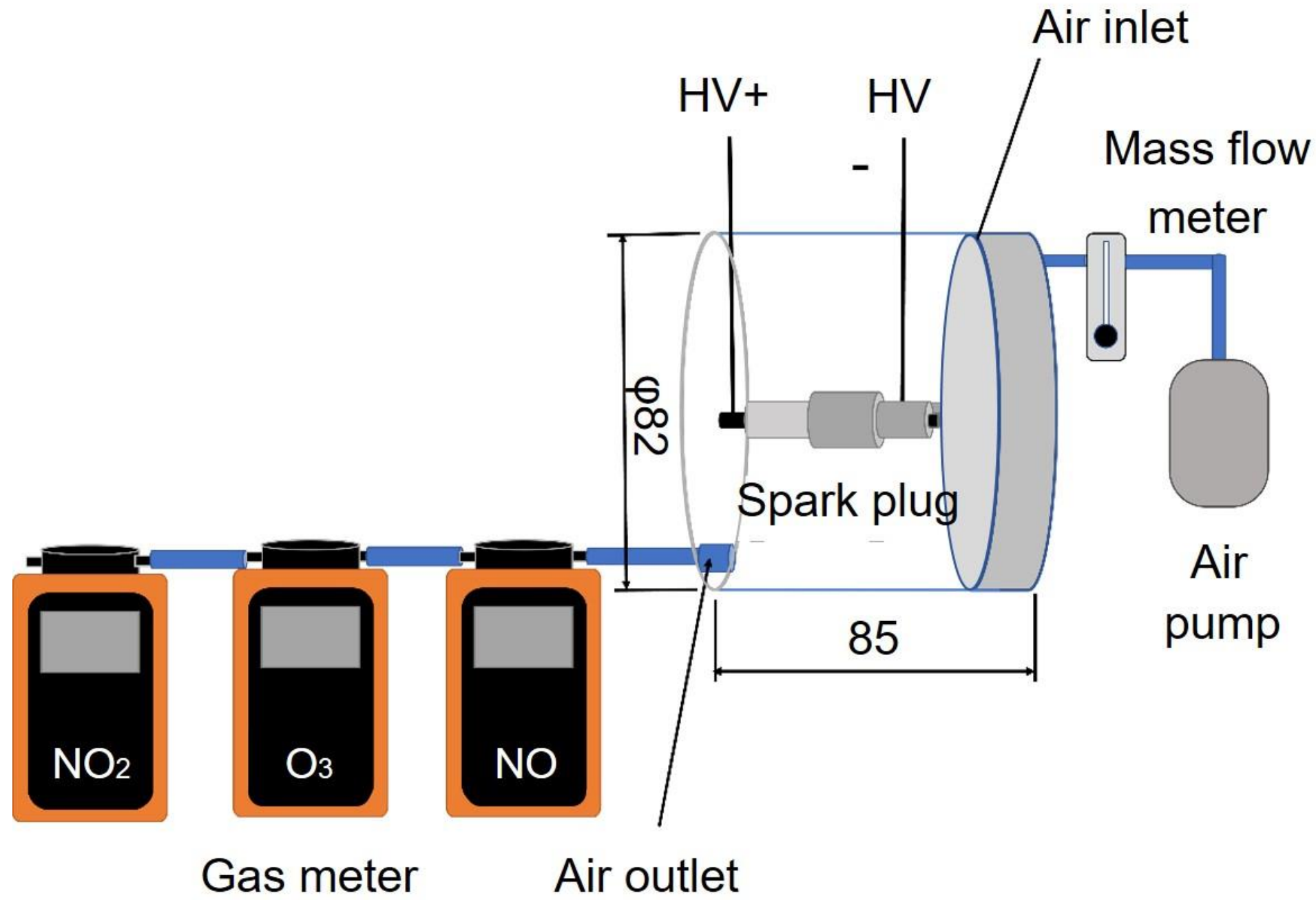

Fig. 3. Schematic of the $NO_x$ and $O_3$ gas measurement setup.

**Rotational and Electrode Temperature Estimation**

The rotational temperature ($T_{rot}$) was used as an approximation of the gas temperature. Under atmospheric pressure discharge conditions, rapid rotational relaxation due to frequent molecular collisions allows rotational levels to reach local thermal equilibrium with translational motion, making $T_{rot}$ a reasonable representation of gas temperature [33,34]. The rotational temperature was obtained by spectral fitting using Specair [33] with optical emission collected by a spectrometer (Acton SP2500, Teledyne Princeton Instruments, United States).

The electrode temperature was measured using an infrared thermal camera (FLIR ONE Edge Pro, Teledyne FLIR LLC, United States) positioned 15 cm above the electrodes to evaluate the thermal accumulation during discharge.

## III. Results

### A. Results of Pulse Rise Time Adjustment

This section presents the post-ignition waveforms obtained using different pulse rise time adjustment methods. The waveforms were recorded under stable discharge conditions and averaged over 1000 pulses to suppress arc-induced fluctuations and reveal consistent trends.

Fig. 4(a) shows the high voltage pulse generated by the IES circuit when triggering the spark plug ($R_g$ = 90 Ω, DC input = 11.1 V). The IES produces sharp pulses at 5 kHz followed by LC oscillations arising from equivalent inductance and parasitic capacitances. Because the oscillation amplitude is far below the breakdown voltage, subsequent analyses focus on the leading pulse.

Fig. 4(b) shows results obtained by varying the external gate resistance $R_g$. The rise time increases slightly with increasing $R_g$. When $R_g$ = 1500 Ω, excessive limitation of gate current prevents reliable MOSFET turn on and spark ignition; therefore, $R_g$ = 30–1000 Ω was adopted. The main waveform differences occur before t = 0 μs, whereas post-ignition rising slopes are nearly identical, indicating that $R_g$ primarily affects turn on delay rather than the voltage rising phase.

Fig. 4(c) presents rise time adjustment using an external drain–source capacitor $C_{ds}$. The rise time first decreases and then increases with increasing $C_{ds}$. When $C_{ds}$ reached 1 μF, ignition was not achievable even at 16 V, so $C_{ds}$ was limited to 10 pF–100 nF. Contrary to simple capacitive charging expectations, a nonlinear behavior is observed. For $C_{ds} \geq 1$ nF, RLC oscillations caused by transformer leakage inductance and MOSFET parasitic capacitances appear after the voltage peak. At $C_{ds}$ = 1–10 nF, the initial oscillatory response accelerates the primary voltage rise, shortening the measured rise time, whereas at 100 nF, the lower resonance frequency weakens this effect, leading to a longer rise time. Thus, $C_{ds}$ -based adjustment suffers from oscillation-induced nonlinearity and poor controllability.

Fig. 4(d) shows results obtained using a parallel capacitor on the transformer secondary side. As $C_{HV}$ increases from 0 to 33 pF, the rise time increases monotonically from 355 ns to 570 ns. Although pulse amplitude decreases, the DC input voltage was adjusted to maintain comparable peak values. Larger $C_{HV}$ causes excessive attenuation and insignificant further rise-time change; therefore, 0–33 pF was selected.

Overall, adding a parallel capacitor ($C_{HV}$) on the transformer secondary provides the most stable and predictable rise-time extension in the present IES architecture. However, this approach inherently couples rise time and pulse width, with longer rise times accompanied by wider pulses. Since both shorter rise time and longer pulse width may promote NO generation [19,22,23], the effects of these coupled waveforms require further investigation.

(a)

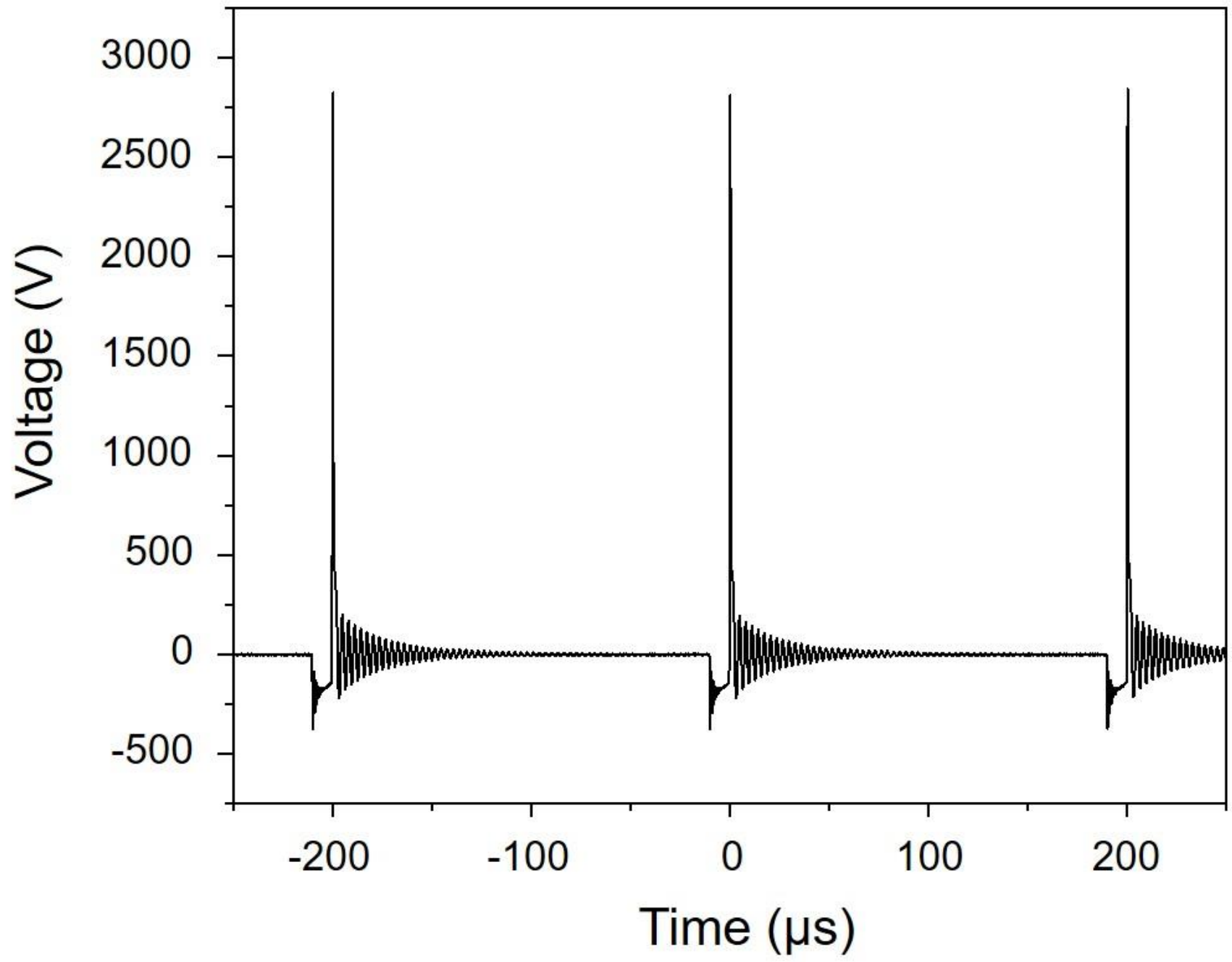

(b)

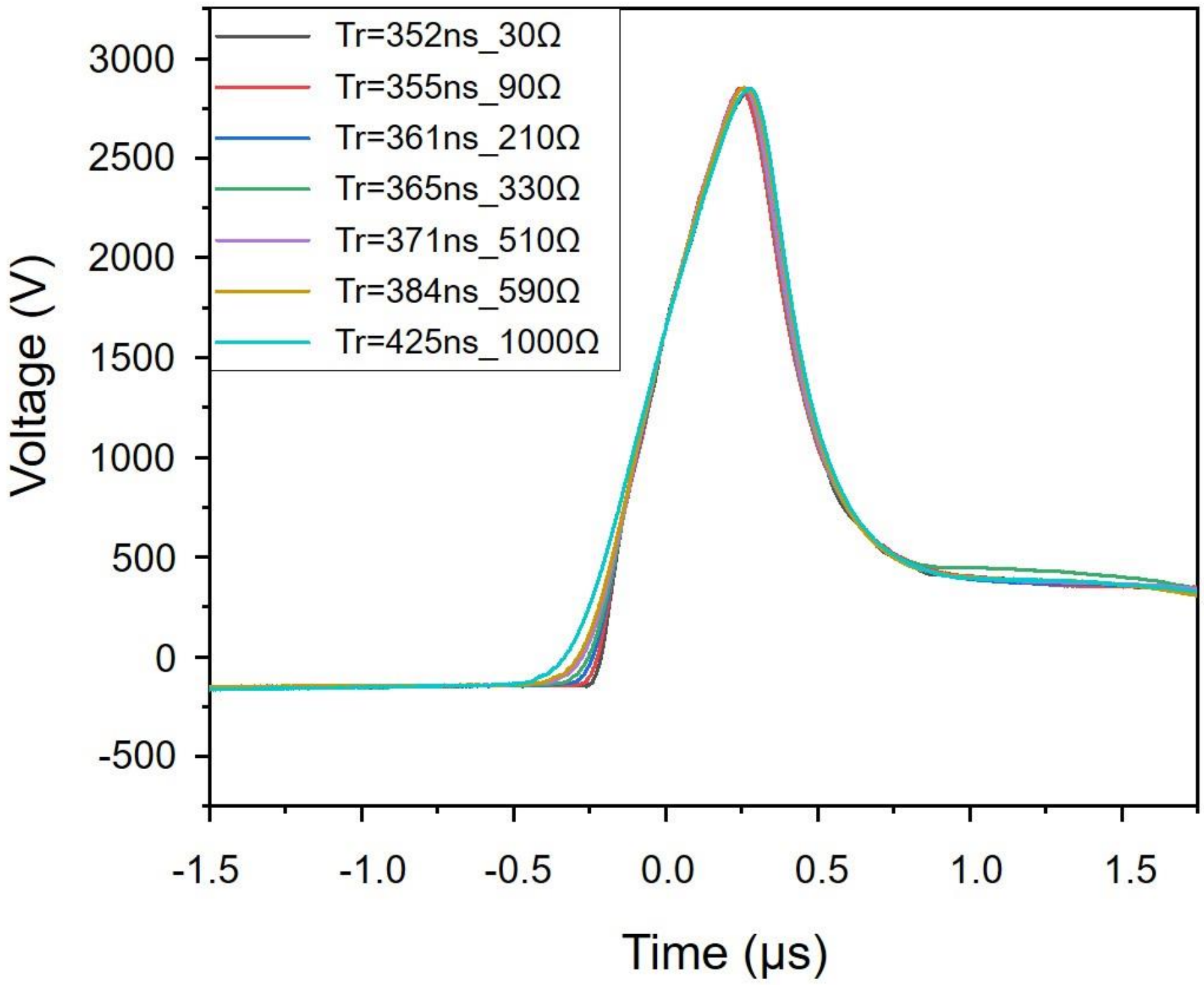
Tr=352ns_30Ω
Tr=355ns_90Ω
Tr=361ns_210Ω
Tr=365ns_330Ω
Tr=371ns_510Ω
Tr=384ns_590Ω
Tr=425ns_1000Ω
Voltage (V)
Time (μs)
3000
2500
2000
1500
1000
500
0
-500
-1.5
-1.0
-0.5
0.0
0.5
1.0
1.5

(c)

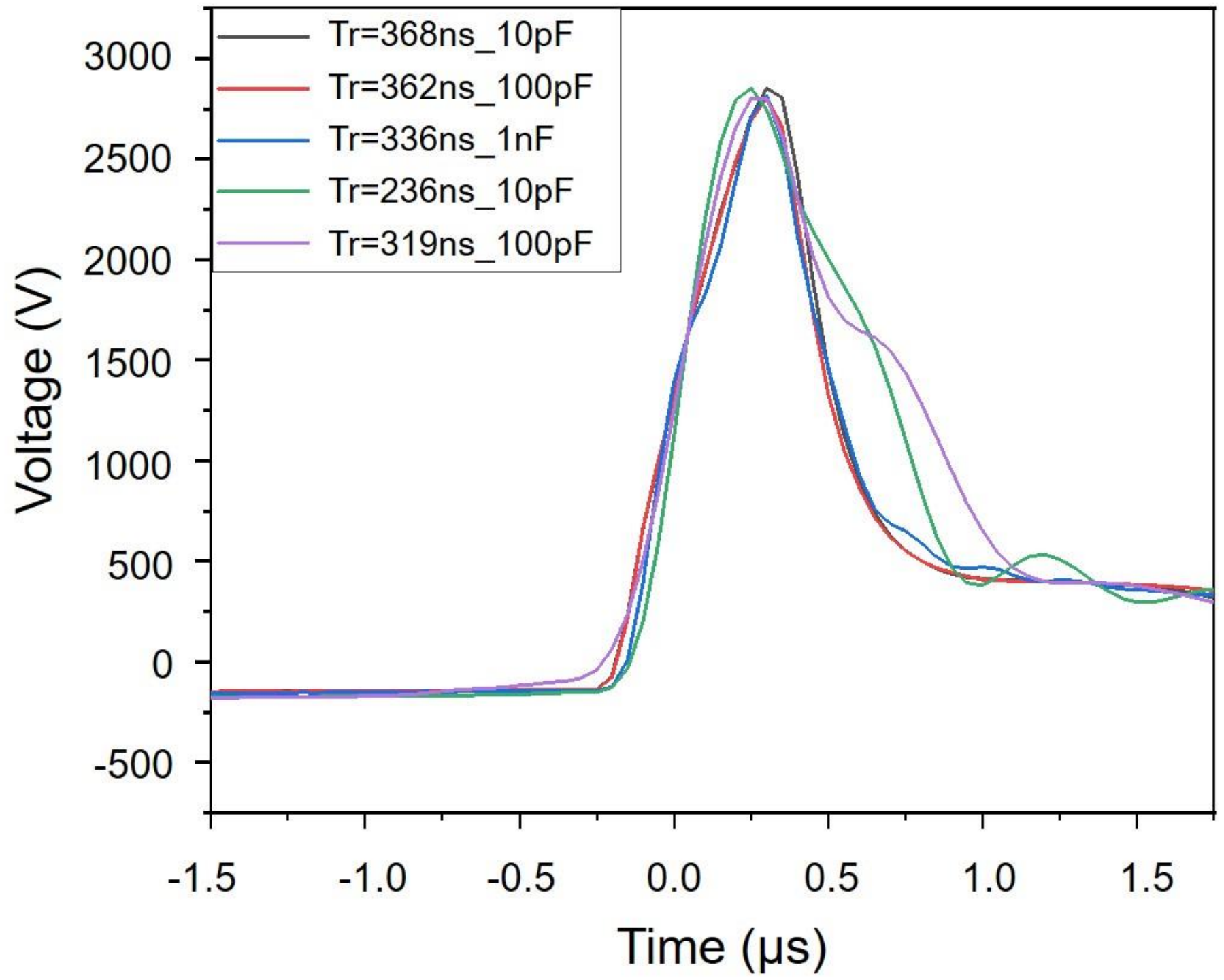
Tr=368ns_10pF
Tr=362ns_100pF
Tr=336ns_1nF
Tr=236ns_10pF
Tr=319ns_100pF
Voltage (V)
Time (μs)
3000
2500
2000
1500
1000
500
0
-500
-1.5
-1.0
-0.5
0.0
0.5
1.0
1.5

(d)

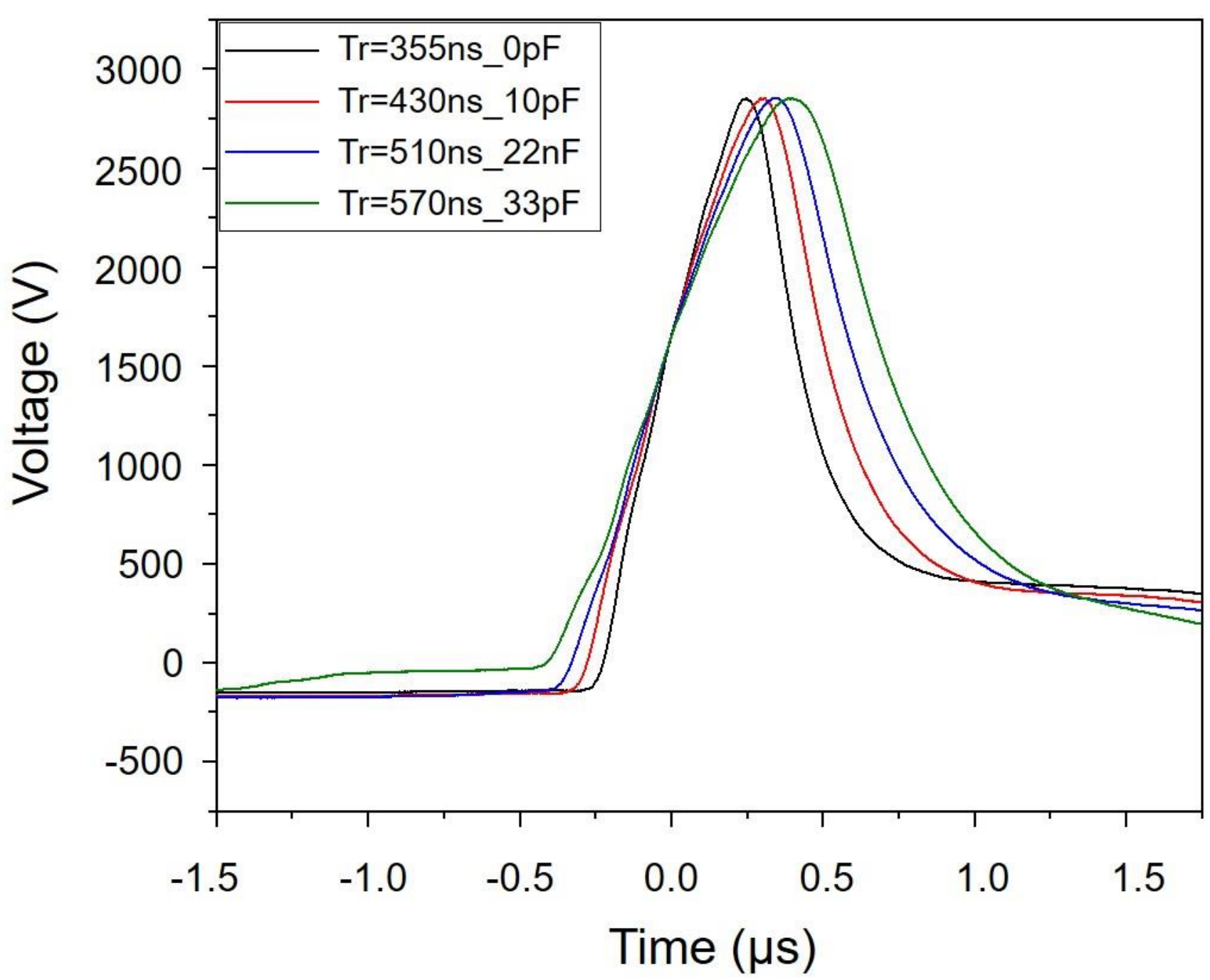

Fig. 4. (a) Basic IES output waveform. Rise time change results for (b) $R_g$, (c) $C_{ds}$, and (d) $C_{HV}$.

## B. Analysis of NO Generation by Spark Discharge under Different Rise Time Conditions

### Voltage and Current Waveforms

Fig. 5 shows the voltage and current waveforms of spark plug discharges in ambient air for different pulse rise times (Tr). The waveforms, labeled A–D in order of increasing Tr, were averaged over 1000 pulses. The pulse repetition rate was fixed at 5 kHz, and the breakdown voltage was maintained at 2.85 ± 0.2 kV. When the voltage reaches the breakdown level, the current rises abruptly, forming a pronounced current peak that indicates gap breakdown and arc formation. As the voltage decreases, the discharge transitions to a glow mode. Detailed waveform parameters are listed in Table 1. The rise time is defined as the 10–90% rise time, and the pulse width is characterized by the full width at half maximum (FWHM).

The waveform with Tr = 570 ns exhibits the highest current peak, and the peak current generally increases with increasing rise time, indicating higher discharge energy for longer rise time waveforms.

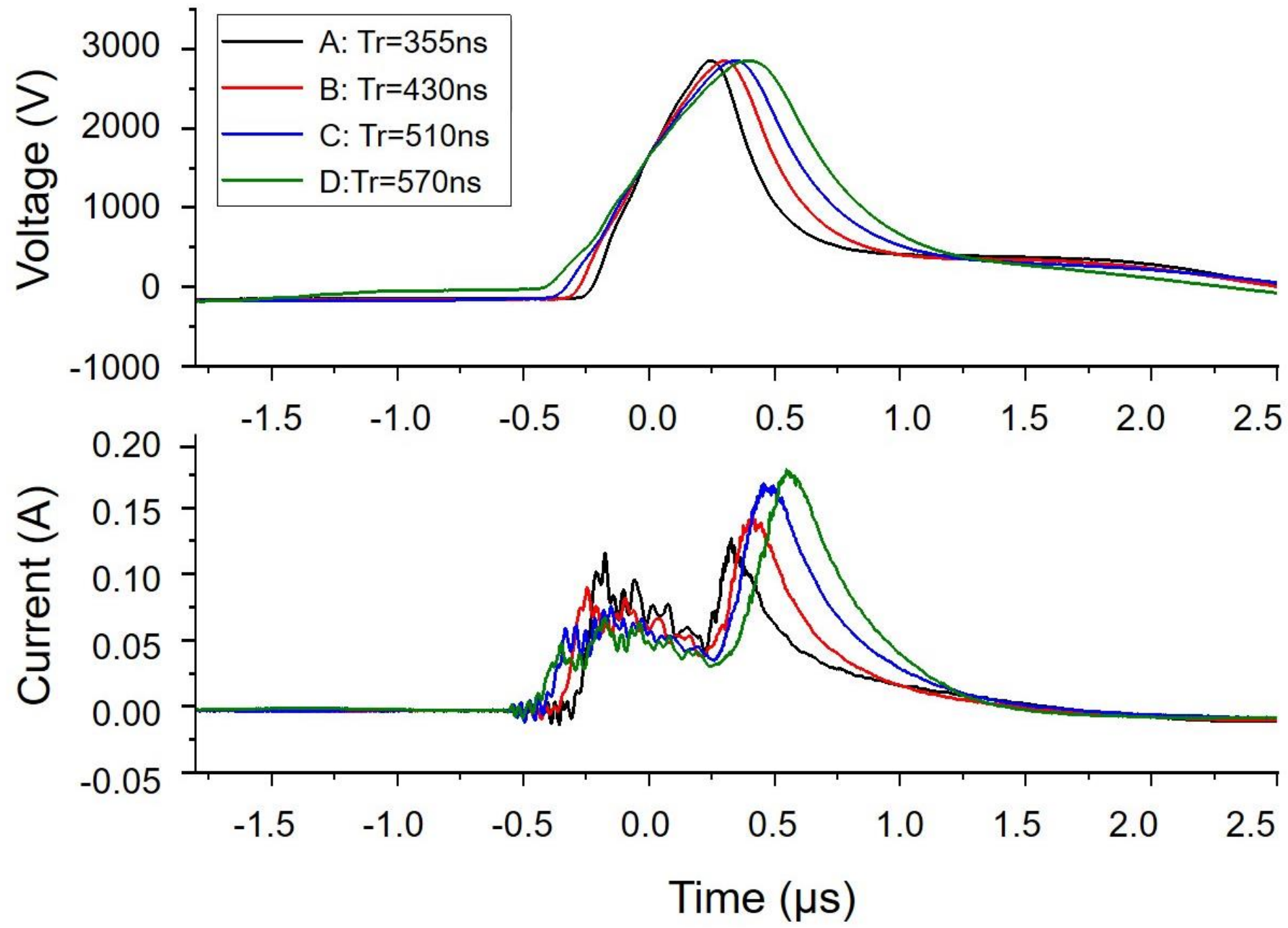

Fig. 5. Voltage and current waveforms under different rise time conditions.

Table 1. Detailed parameters of the four pulse waveforms.

| | **Tr** | $\mathbf{V_{DC}}$ | $\mathbf{C_{HV}}$ | **FWHM** |
|---|---|---|---|---|
| A | 355 ns | 11.1 V | 0 pf | 462.2 ns |
| B | 430 ns | 12.2 V | 10 pf | 570.2 ns |
| C | 510 ns | 13.2 V | 22 pf | 690.8 ns |
| D | 570 ns | 15.5 V | 33 pf | 771.3 ns |

**Optical Emission Spectroscopy Results**

Emissions from the $N_2$ second positive system (~316 nm, $C^3\Pi_u \rightarrow B^3\Pi_g$), the $N_2^+$ first negative system (~391 nm, $B^2\Sigma^+_u \rightarrow X^2\Sigma^+_g$), and the NOγ band (~236 nm, $A^2\Sigma^+ - X^2\Pi$) were selected to evaluate the effects of the four waveforms on plasma excitation and reactive species generation. The $N_2$ second positive emission reflects electron-impact excitation of nitrogen molecules, with vibrationally excited $N_2^*$ regarded as an important precursor for NO formation [35]. The $N_2^+$ first negative emission originates from direct electron-impact ionization and indicates the plasma ionization degree, while the NOγ emission represents the formation of excited NO(A) species and serves as a reference for NO production [36].

As shown in Fig. 6(a)–(c), the emission intensities of $N_2^*$, $N_2^+$, and NOγ all increase with increasing rise time. This trend suggest that longer rise times are likely associated with higher electron energies, leading to enhanced electron-impact excitation and ionization processes and thereby promoting reactions related to NO generation.

(a)

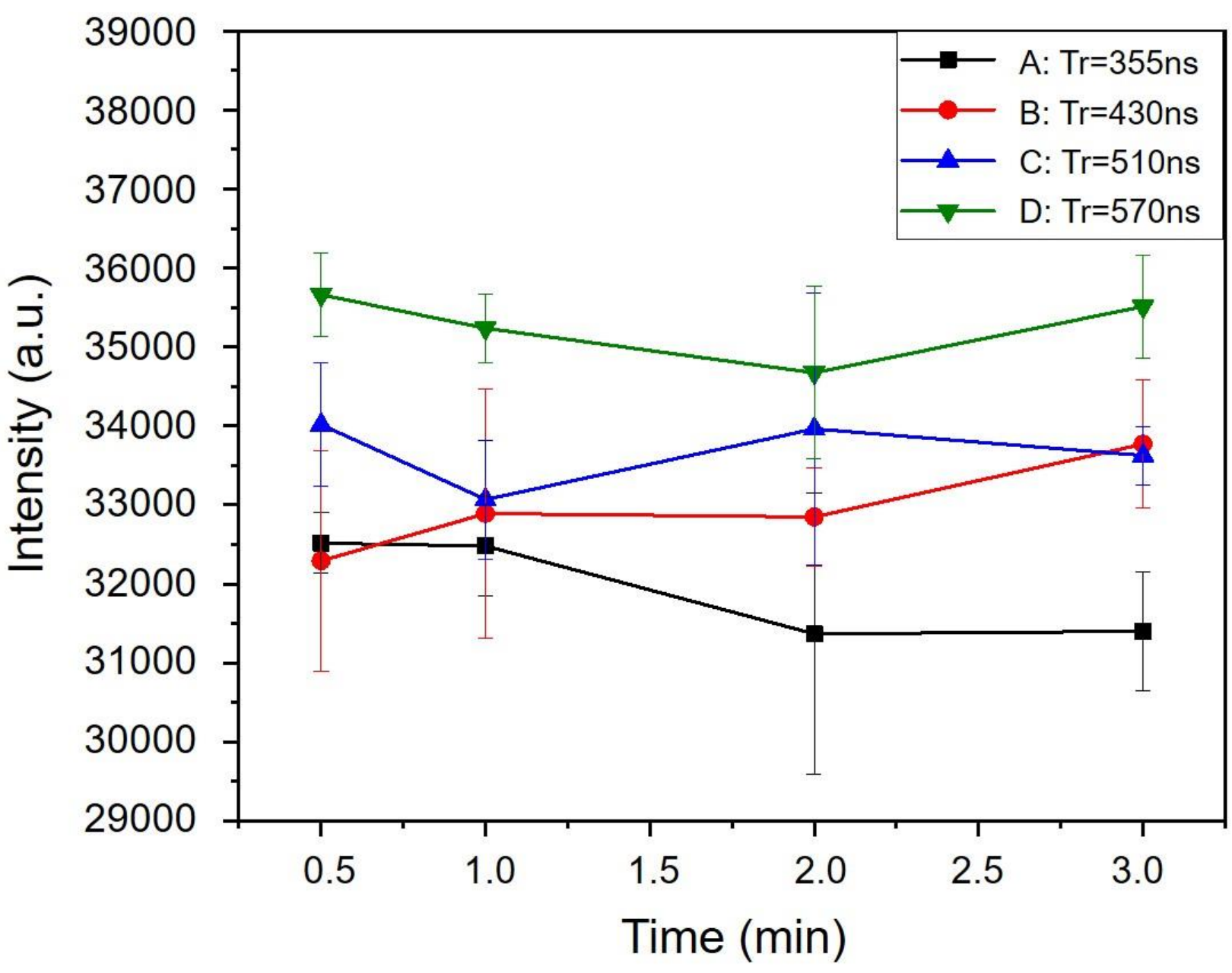

(b)

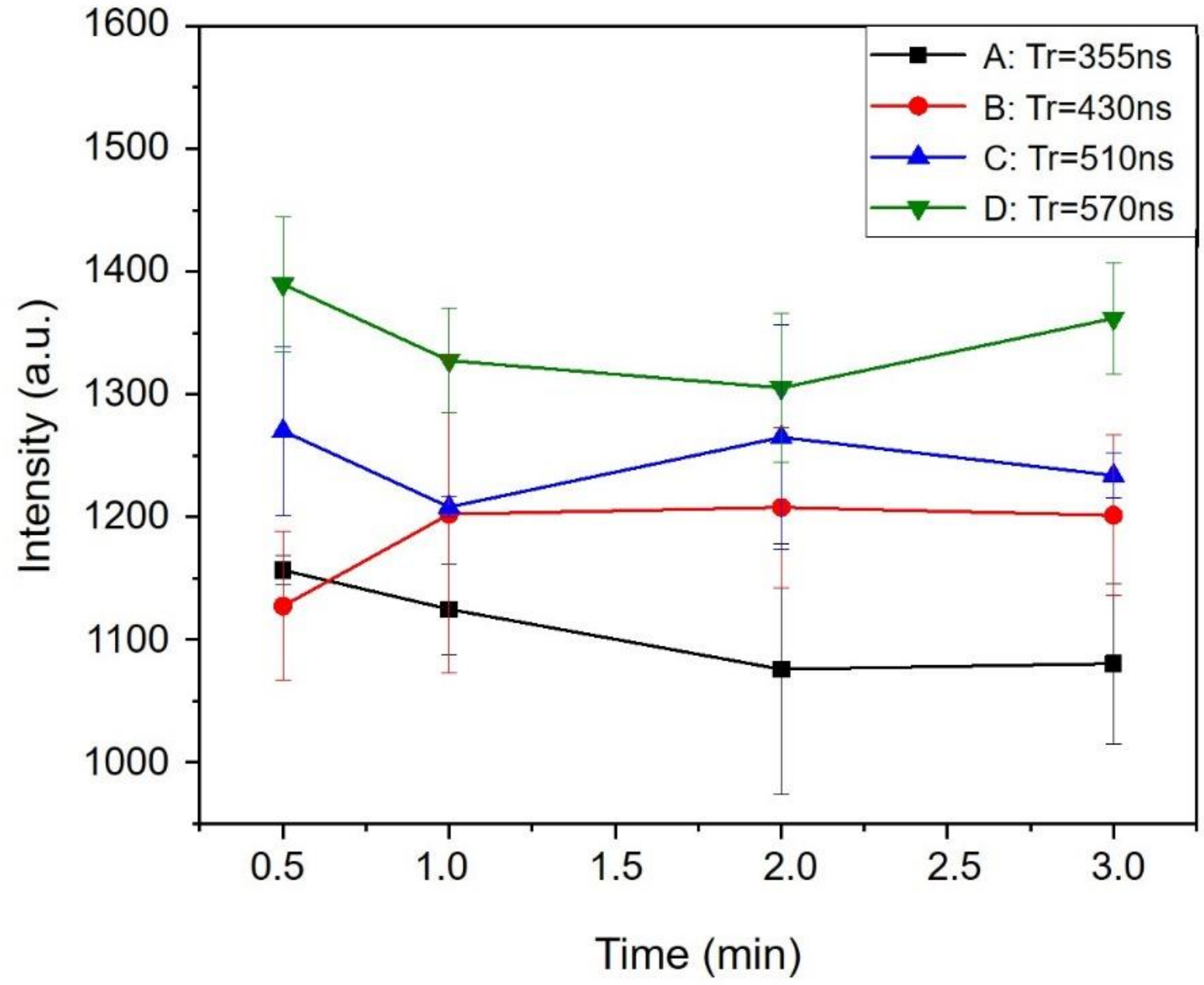

(c)

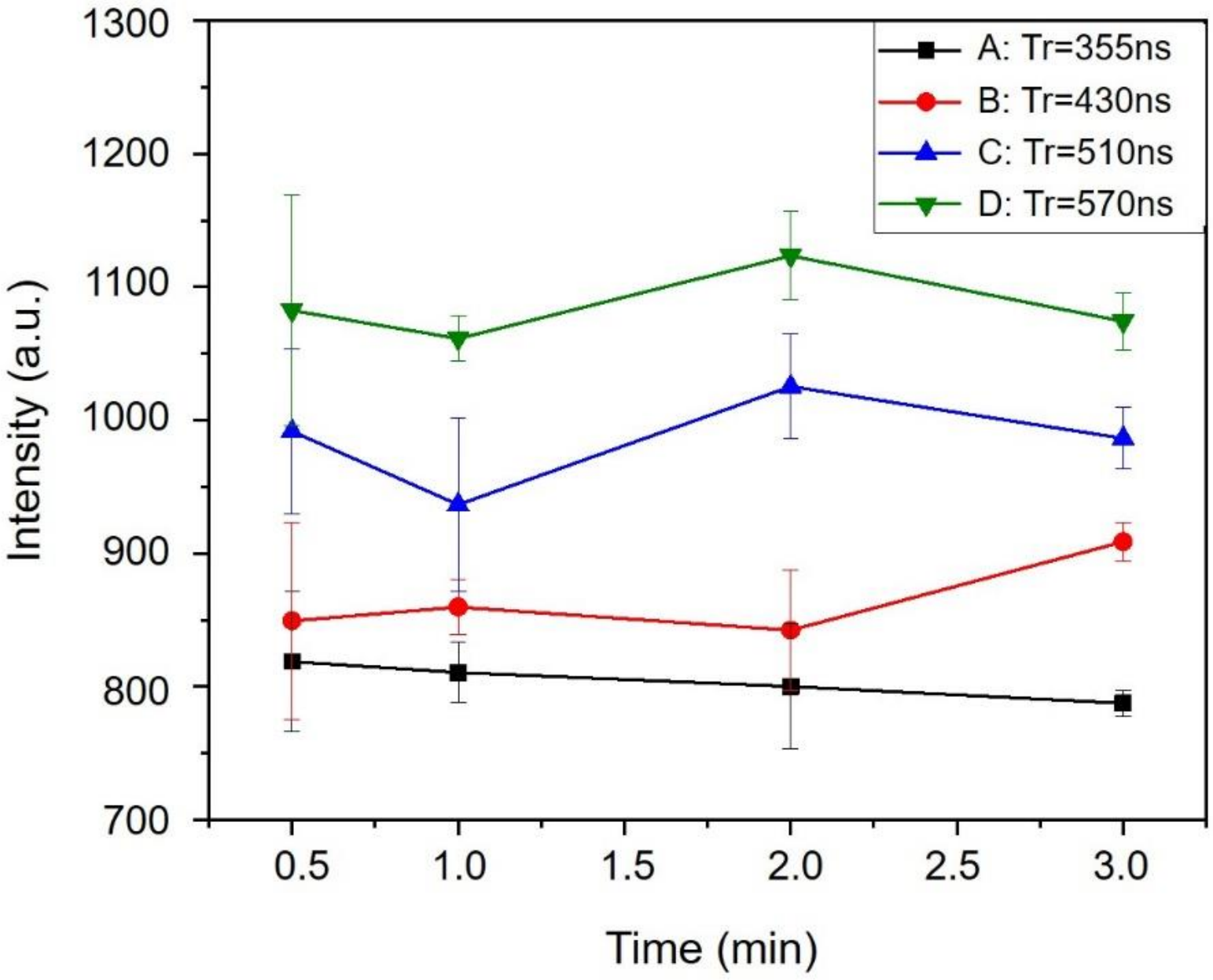

Fig. 6. Optical emission intensities of different waveforms: (a) $N_2^*$ at 316 nm, (b) $N_2^+$ at 391 nm, and (c) NOγ at 236 nm.

**$NO_x$ and $O_3$ Gas Measurement Results**

$NO_x$ and $O_3$ analyzers were used to quantify post-discharge gas concentrations. As shown in Fig. 7(a), the NO concentration increases with discharge time under all rise time conditions, with the highest NO level observed for Tr = 570 ns. Overall, the NO generation rate increases with increasing rise time. Similar trends are observed for $O_3$ and $NO_2$, as shown in Fig. 7(b) and Fig. 7(c).

A Tukey HSD test performed on the NO concentrations shows statistically significant differences only between Tr = 355 ns and Tr = 510 ns or 570 ns, whereas no significant differences are observed among Tr = 430 ns, 510 ns, and 570 ns. This behavior is attributed to the accumulation of NO over the measurement period, which averages out intermediate differences. As a result, discernible separation remains primarily between the shortest rise time condition and the longer rise time conditions. Despite the lack of clear distinction among the intermediate groups, the overall results demonstrate that the rise time adjusted waveforms produced in this study do influence NO generation.

(a)

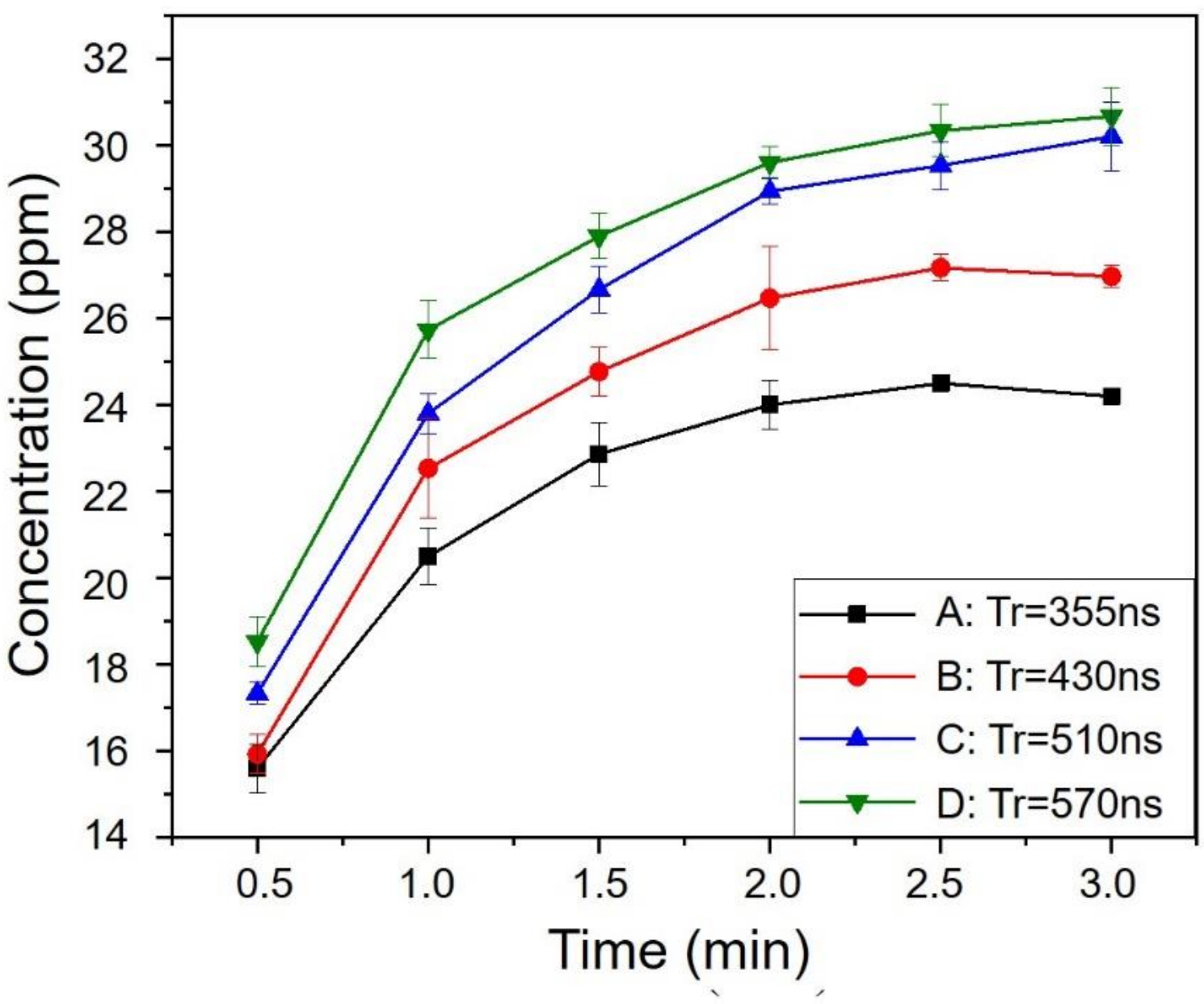

(b)

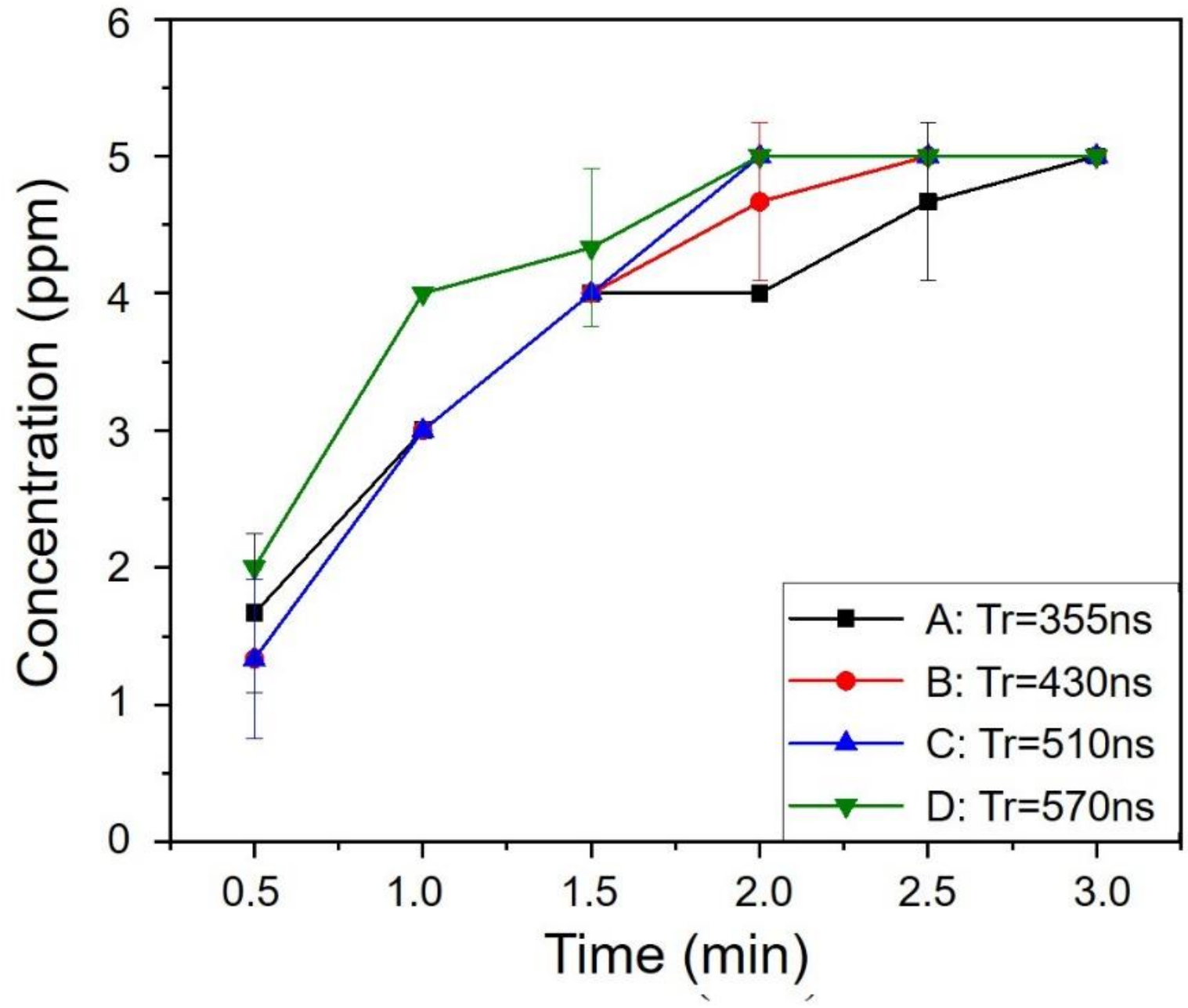

(c)

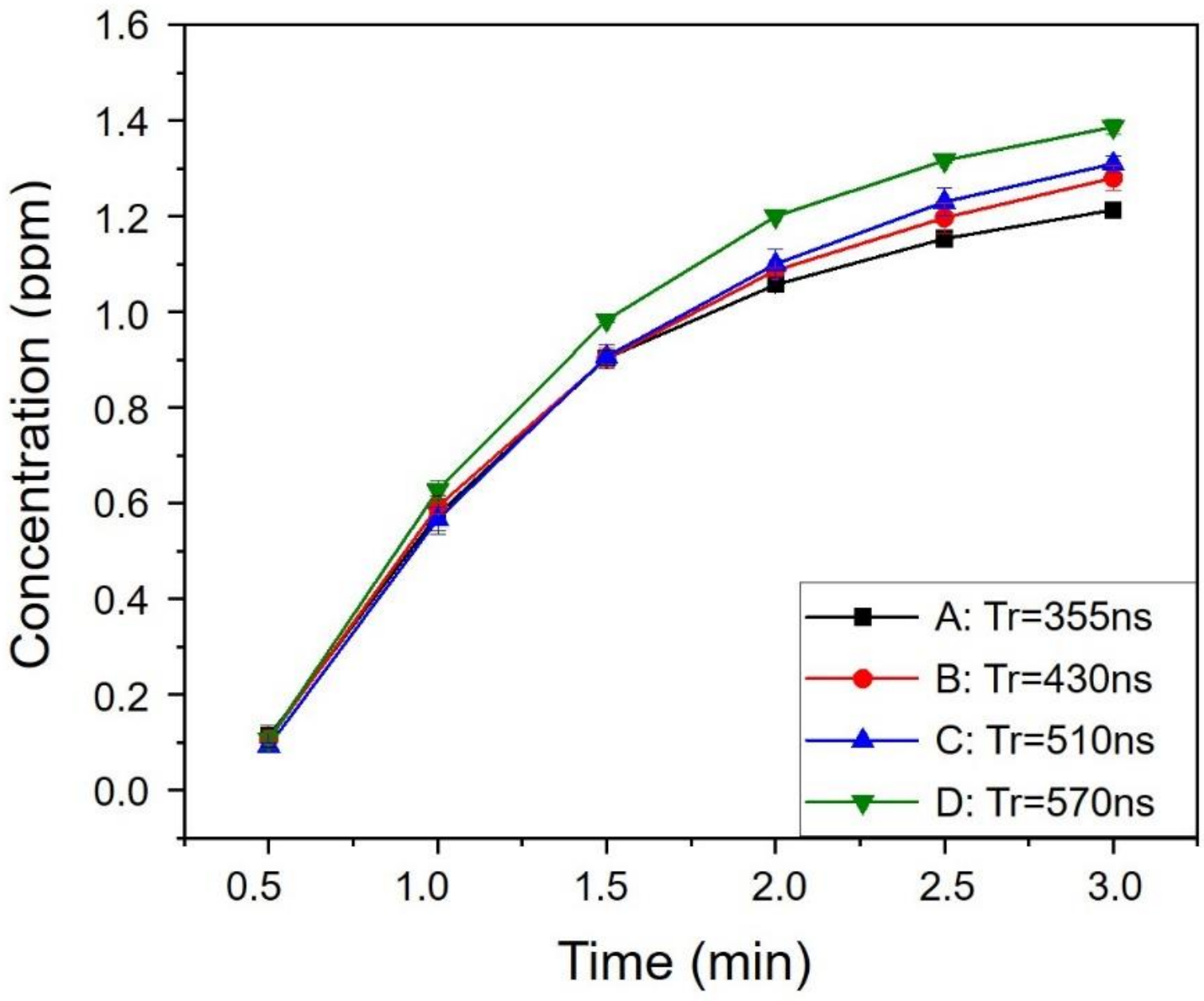

Fig. 7. Concentrations of (a) NO, (b) $O_3$, and (c) $NO_2$ under different waveform conditions.

**Electron Excitation Temperature Estimation Results**

The Ar I emission lines at 750.387 nm and 426.628 nm were used to estimate the electron excitation temperature $T_e$, with the 751.465 nm line additionally employed to verify trend consistency. The measured intensities and atomic parameters from the NIST Atomic Spectra Database (Table 2) were substituted into Eq. (8) [31].

$$\frac{I_{nm}}{I_{qp}} = \frac{g_n A_{nm} \lambda_{qp}}{g_q A_{qp} \lambda_{nm}} \exp\left(-\frac{E_{nm} - E_{qp}}{K_B T_e}\right) \tag{8}$$

Fig. 8 shows the electron excitation temperatures derived from the Ar I 426.628 / 750.387 nm intensity ratio. The highest $T_e$ is obtained for Tr = 570 ns at all measurement times, whereas the lowest values occur for Tr = 355 ns, consistent with the trends observed in OES intensities and NO production. Similar trends are obtained using the Ar I 426.628/751.465 nm line pair (Appendix Fig. a), confirming the reliability of the estimation.

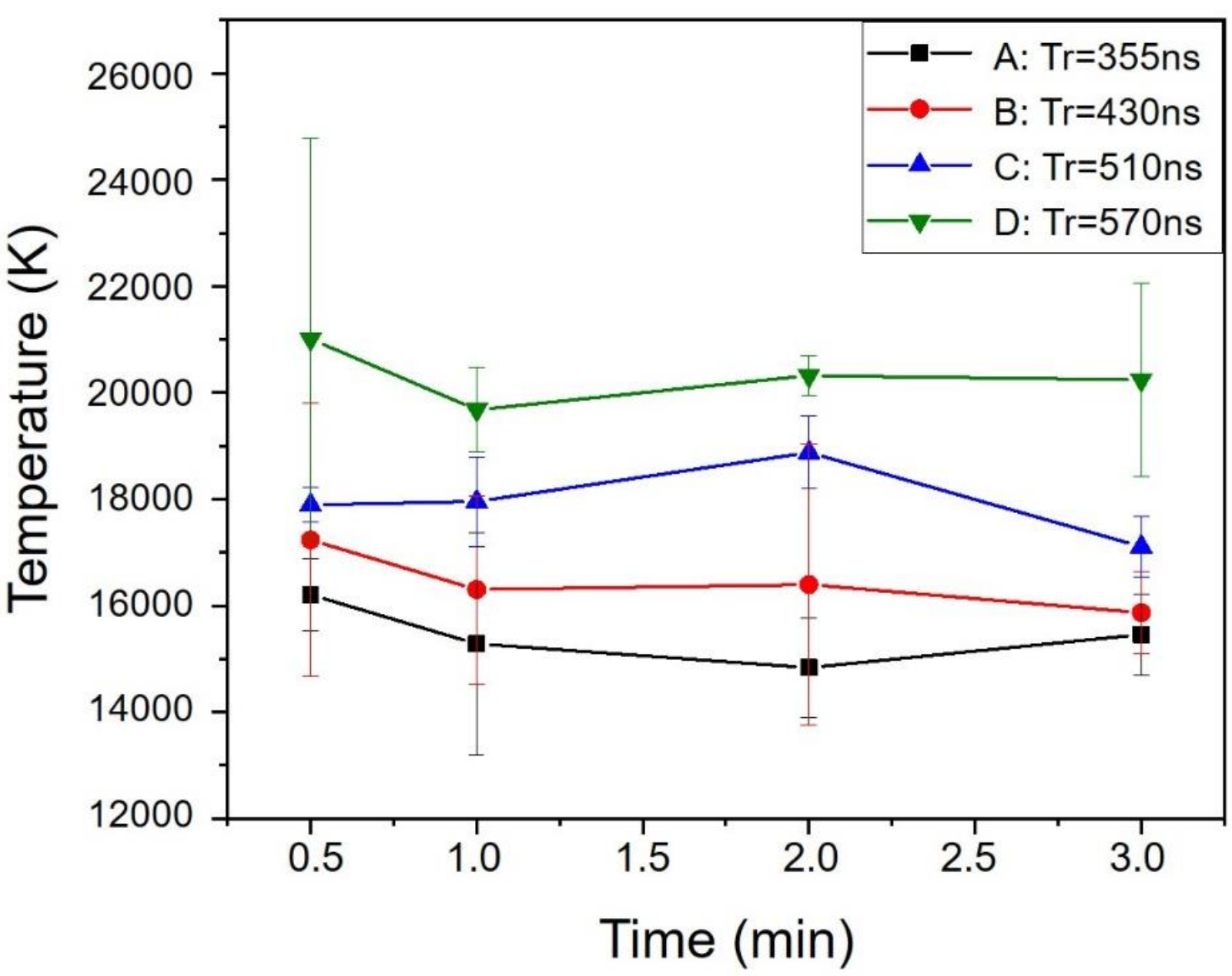

Fig. 8. Electron excitation temperature estimated from the Ar I (426.628 nm / 750.387 nm) line intensity ratio.

Table 2. Parameters used for electron excitation temperature calculation based on the line intensity method.

| λ(nm) | Transition | $A_{nm}$ ($s^{-1}$) | $g_i$ | $g_n$ | $E_n$ (ev) |
|---|---|---|---|---|---|
| 426.6286 | 4p – 4s | 3.10E+05 | 3 | 5 | 117183.5901 |
| 750.3869 | 4p – 4s | 4.50E+07 | 3 | 1 | 108722.6194 |
| 751.4652 | 4p – 4s | 4.00E+07 | 3 | 1 | 107054.272 |

Overall, the electron excitation temperature increases with increasing rise time. However, because rise time and pulse width are inherently coupled in the present waveforms, the dominant mechanism remains unclear. Therefore, FFT analysis and plasma absorbed power calculations are performed to further elucidate the origin of this trend.

**Fast Fourier Transform Analysis Results**

Because electron response to an external electric field exhibits characteristic frequencies, higher frequency field components are more effective in driving electron acceleration and promoting excitation, dissociation, and reactive species generation. Accordingly, FFT analysis was performed on the four pulse waveforms to examine the relationship between rise time and electron excitation temperature.

As shown in Fig. 9, the waveform with Tr = 355 ns exhibits only a slight increase in high frequency components , while the overall spectral profiles of all four waveforms are highly similar. Therefore, the electron acceleration capability in the plasma resulting from the different rise time conditions can be regarded as comparable. These results suggest that the observed differences in electron excitation temperature are more likely governed by other discharge parameters rather than by the pulse rise time alone.

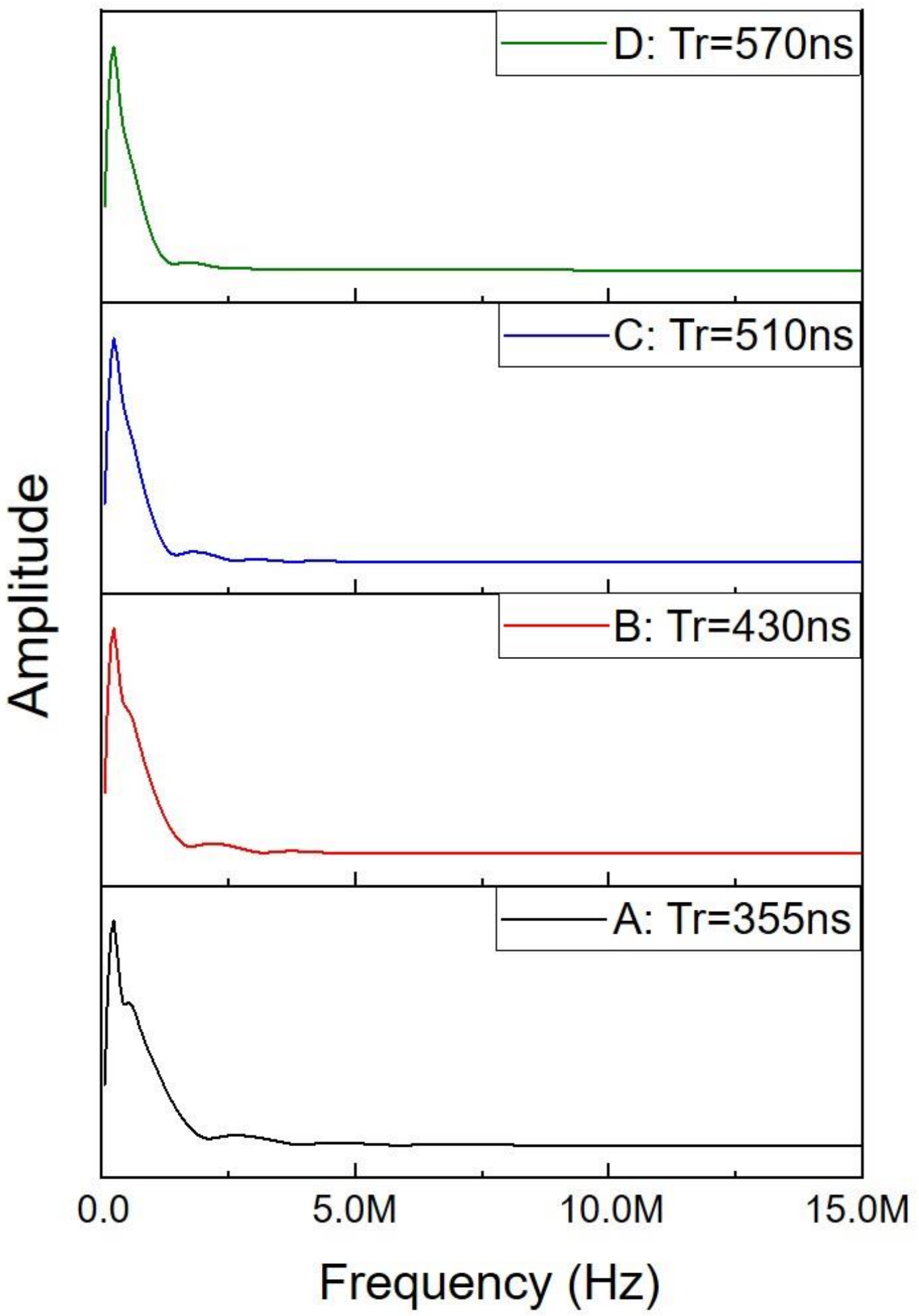

Fig. 9. Fast Fourier transform (FFT) spectra of the different voltage waveforms.

**Plasma Absorbed Power Calculation Results**

Following the method reported in [32], the plasma power delivered to the spark plug gap was estimated as:

$$P_{gap} = \int_{t0}^{t1} (V(t) * I(t) - I(t)^2 R_p)\, dt * PRR \quad (12)$$

The equivalent internal resistance of the spark plug was measured to be approximately 1.77 kΩ using an LCR meter. The voltage and current waveforms recorded by the oscilloscope, together with a pulse repetition rate of 5 kHz, were substituted into the calculation. The resulting absorbed plasma power is shown in Fig. 10.

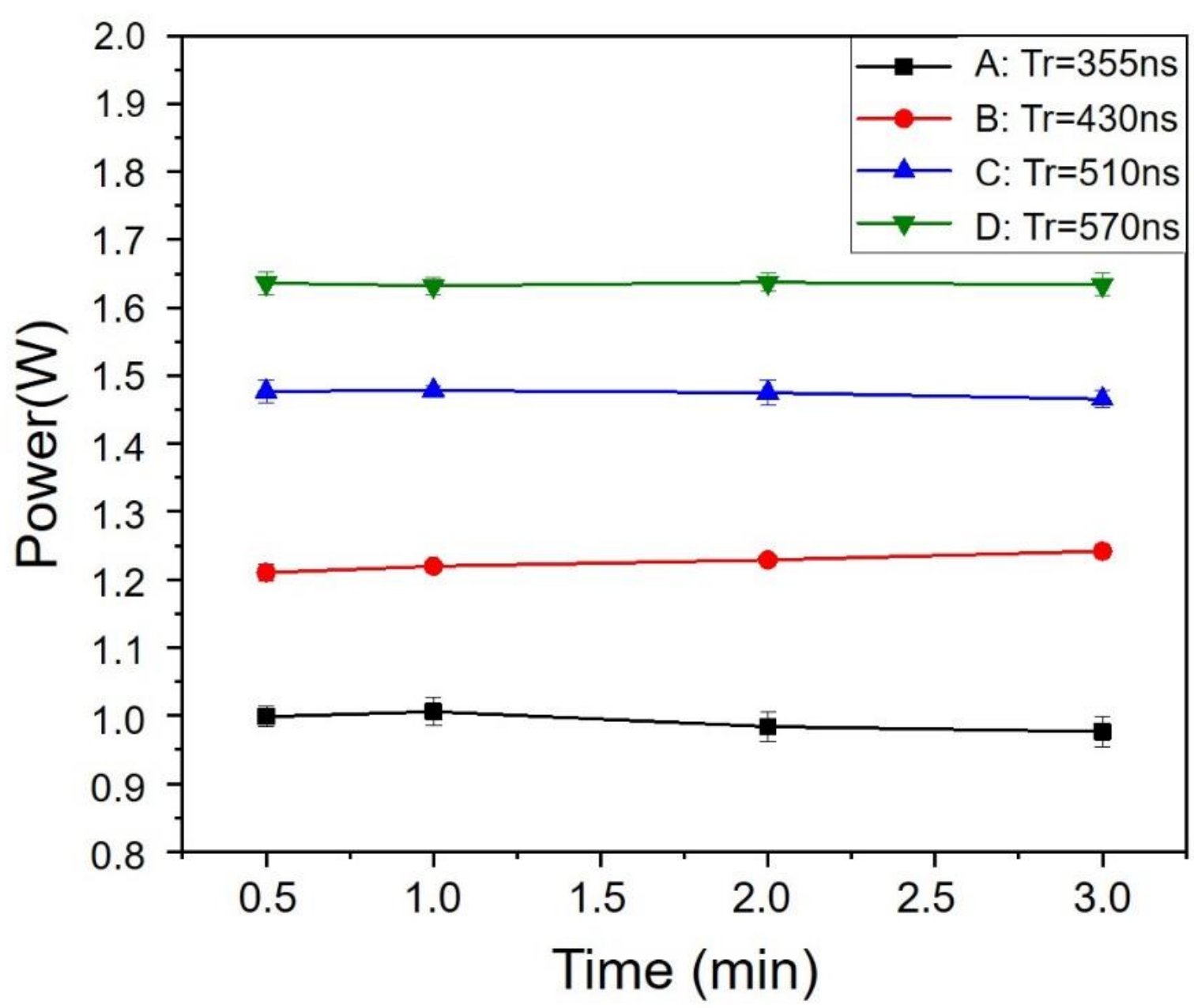

Fig. 10. Plasma absorbed power for different waveforms.

All four waveforms exhibit stable power input over time. The absorbed plasma power increases with rise time, with the highest value at Tr = 570 ns and the lowest at Tr = 355 ns, consistent with the measured electron excitation temperatures. Because a longer pulse width extends the electric-field action and allows electrons to absorb more energy within a single pulse, increased excitation and collision probabilities are expected, as also reported previously [19].

Combining the absorbed power, electron excitation temperature, and FFT results shows that although minor differences appear at high frequencies, the overall spectral profiles are similar among the waveforms. Thus, electron acceleration capability is likely comparable for different rise times, whereas the increased effective energy input associated with pulse width extension provides a more plausible explanation for the observed rise in electron excitation temperature. Accordingly, within the compact IES architecture, electron energy and excitation temperature are primarily governed by pulse duration rather than by rise time.

### C. Thermal Effects on NO Generation

Rotational temperature was estimated by Specair [33] fitting of the $N_2$ second positive system (334–337.5 nm) to assess the influence of gas heating on NO generation. As shown in Fig. 11, all four rise time conditions yield rotational temperatures of approximately 600–650 K, with no clear rise time dependence. Because the spike-like pulses used in this study are expected to produce rapid temporal variations within a single discharge cycle, the absence of a discernible trend is attributed to the use of time-integrated OES rather than time-resolved, synchronized measurements, which may average out transient differences.

To further evaluate thermal effects, electrode surface temperature was measured using an infrared thermal camera (Fig. 12). The electrode temperature increases with increasing rise time, consistent with trends in electron excitation temperature and plasma absorbed power, indicating greater effective energy input and thermal accumulation for longer rise time waveforms. Although the thermal decomposition rates of $O_3$ and $NO_2$ are temperature dependent [12,13], the maximum electrode temperature difference among the four cases is only about 9 ℃, and the absolute temperature remains below approximately 50 ℃, far below that required to activate the Zeldovich mechanism [9–13]. Therefore, thermal effects are expected to play a minor role, and the observed trends are more likely dominated by electron-impact excitation and ionization processes induced by the adjusted pulse waveforms.

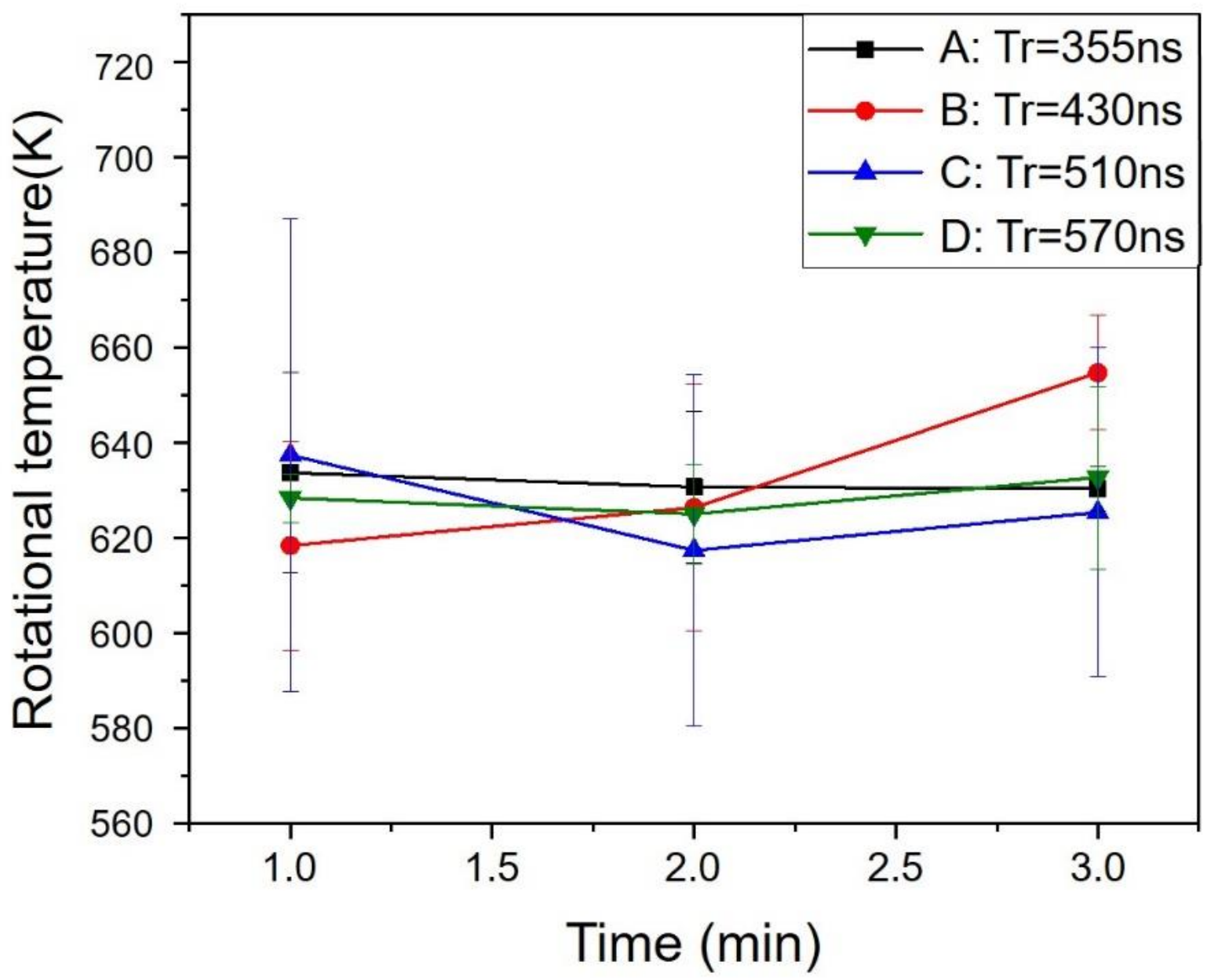

Fig. 11. Rotational temperature fitting results for different waveforms.

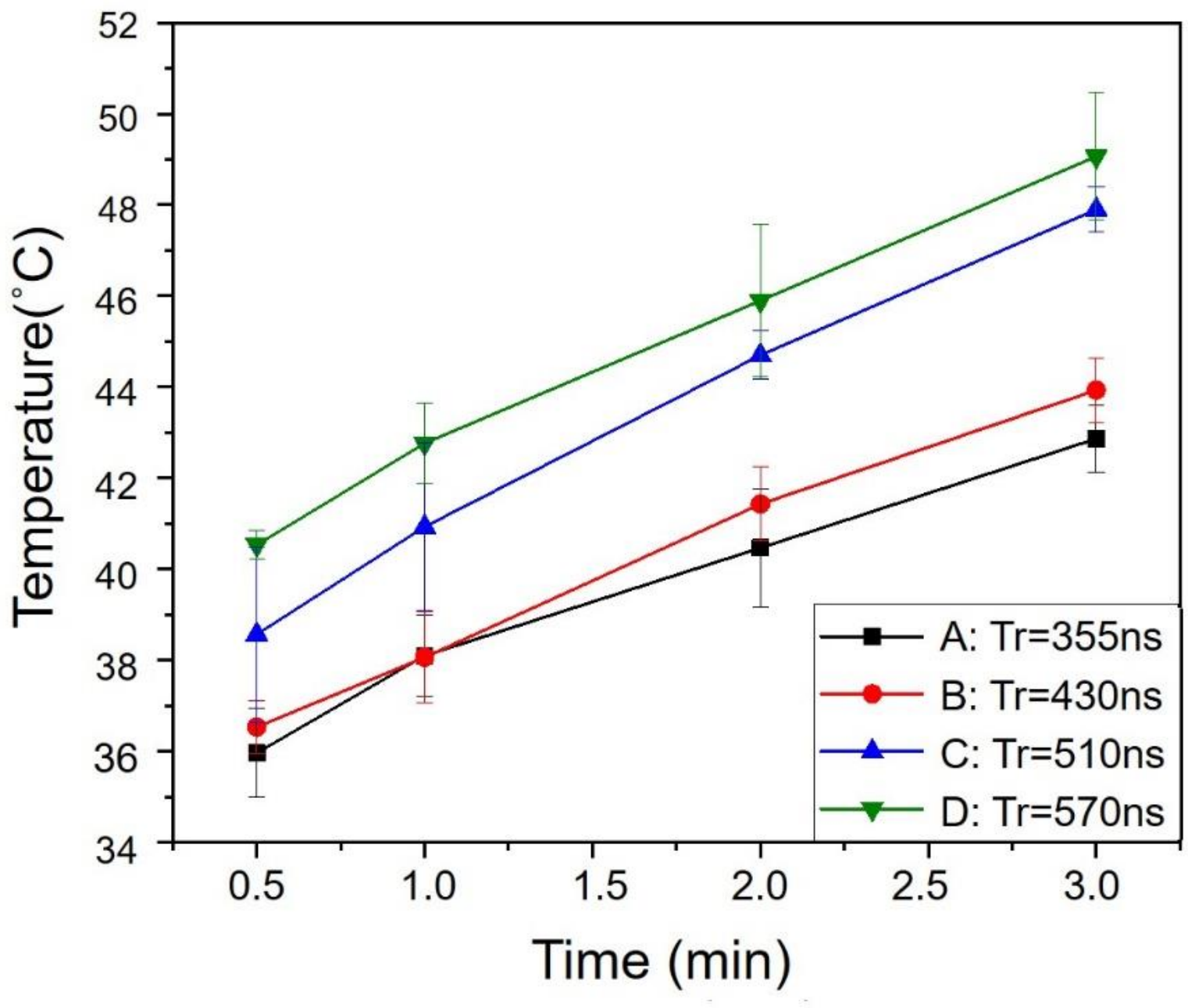

Fig. 12. Electrode temperature measurements for different waveforms.

## IV. Discussion

This study compares three approaches for pulse rise time adjustment. The results indicate that adjusting the external gate resistance primarily affects the turn on delay and provides only limited control over the actual rise time, while adding a drain–source capacitor modifies the waveform shape but leads to poor stability and inconsistent rise time behavior. In contrast, introducing a high voltage capacitor in parallel with the transformer secondary enables a stable and predictable extension of the rise time and appears to be the most feasible adjustment approach within a compact IES circuit. Nevertheless, this method inherently couples the rise time with the pulse width, preventing independent control of these parameters and representing a key limitation of rise time adjustment in the present system.

Under these waveform conditions, OES, $NO_x$ and $O_3$ concentration analyses, and electron excitation temperature estimates consistently show that longer rise times are associated with enhanced molecular excitation, ionization, and increased NO production. To interpret the origin of these trends, FFT and plasma absorbed power analyses were further examined. FFT analysis indicates that the spectral distributions of different waveforms are highly similar, suggesting comparable electron acceleration capability. In contrast, plasma absorbed power increases with pulse width and correlates strongly with both NO production and electron excitation temperature. This correspondence suggests that the observed differences are more strongly related to the increase in effective energy input associated with pulse width extension rather than to the rise time itself. Consequently, within the compact IES architecture, rise time cannot be treated as an independent parameter for isolating its direct influence on NO generation.

Thermal effects were further evaluated using rotational and electrode temperature measurements. Although no clear trend is observed in rotational temperature, likely due to limited temporal resolution, electrode temperature increases systematically with rise time, indicating greater thermal accumulation for longer rise time waveforms. Nevertheless, the maximum electrode temperature difference is only about 9 ℃, and the absolute temperature remains below approximately 50 ℃, which is insufficient to activate the Zeldovich mechanism and suggests limited differences in the thermal decomposition of $O_3$ and $NO_2$. Therefore, thermochemical pathways are expected to play a secondary role under the present conditions, and the observed trends are dominated by electron-impact excitation and ionization processes induced by the adjusted rise time pulse waveforms.

## V. Conclusion

This study investigated pulse rise time adjustment in a portable inductive energy storage (IES) circuit and its influence on plasma characteristics and NO generation. The results show that varying the external gate resistance or adding a drain–source capacitor does not provide stable or predictable rise-time control. In contrast, a parallel capacitor on the transformer secondary enables a stable increase in rise time but inherently couples rise time with pulse width, preventing independent waveform control in the present architecture.

Combined analyses of OES, $NO_x$ and $O_3$ concentration measurements, electron excitation temperature, FFT, and plasma absorbed power indicate that although NO-related species concentrations and electron excitation temperature increase with rise time, the dominant factor is not the rise time itself. FFT results show similar spectral distributions among different waveforms, whereas plasma absorbed power correlates strongly with both NO production and electron excitation temperature, demonstrating that the increased energy input associated with pulse width extension governs the observed trends.

Thermal analyses based on rotational and electrode temperatures further suggest that thermochemical effects are limited under the present operating conditions. Overall, within the compact IES system investigated here, pulse width and energy input, rather than rise time, are the primary parameters controlling plasma behavior and NO generation, and the effectiveness of rise time adjustment is inherently constrained by circuit level coupling.

## Appendix

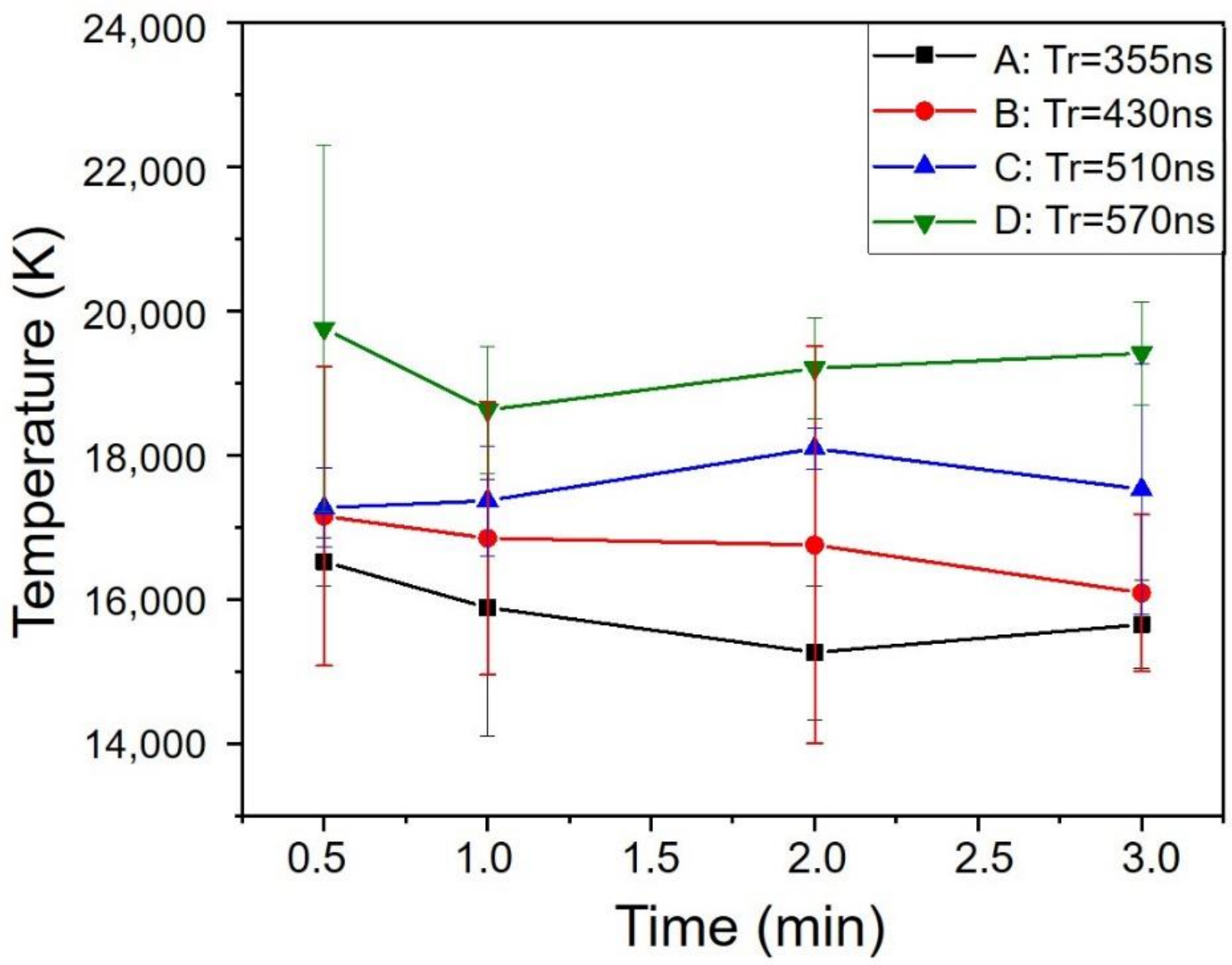

Fig. a. Electron excitation temperature estimated from the Ar I (426.628 nm / 751.465 nm) line intensity ratio.